%% file: main.tex
\documentclass[journal=jpclcd,manuscript=letter]{achemso}
\setkeys{acs}{chaptertitle = true} 

\usepackage[version=3]{mhchem} % Formula subscripts using \ce{}
\usepackage{MnSymbol}
\usepackage{hyperref}
\usepackage{threeparttable}
\usepackage{xspace}
\usepackage{dcolumn}
\usepackage{array}   % for \newcolumntype macro
\hypersetup{
    colorlinks=true,
    linkcolor=blue,
    filecolor=magenta,      
    urlcolor=blue,
    pdftitle={Overleaf Example},
    pdfpagemode=FullScreen,
    }

\input{macros}

\author{Devin M. Mulvey}
\affiliation{Department of Chemistry, St. Bonaventure University, St Bonaventure, New York, 14778, United States of America}
\alsoaffiliation{Department of Chemistry, University of Pittsburgh, Pittsburgh, Pennsylvania, 15260, United States of America}
\author{Kenneth D. Jordan}
\affiliation{Department of Chemistry, University of Pittsburgh, Pittsburgh, Pennsylvania, 15260, United States of America}
\email{jordan@pitt.edu}
\author{Alston J. Misquitta}
\affiliation{Department of Physics and Astronomy, Queen Mary, University of London, London E1 4NS, United Kingdom}

\title[Rational definition of properties of AIMs]
  {Is the atomic quadrupole moment of a carbon atom in graphene zero?: The case for a rational definition of the properties of atoms in a molecule.}

\abbreviations{AIMs, GDMA, BS-ISA}

\begin{document}

\begin{tocentry}
\includegraphics{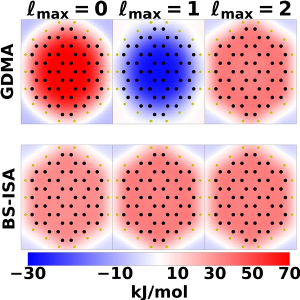}  
\end{tocentry}

\begin{abstract}
  It is generally assumed that the carbon atoms of graphitic samples and their finite analogs have sizable quadrupole moments, with the out-of-plane component ($\QCsph$ in traceless spherical coordinates) being the dominate contribution. 
  However, there is no consensus on what the quantity is for such carbon-based systems and values reported in the literature range from $\QCsph \sim -1.14$ to $+0.79$ a.u. 
  In this work we propose a theoretical framework in which well-defined statements can be made about properties of atoms-in-a-molecule (AIMs) even when these properties are not experimentally observable.
  Using this framework and the distributed multipole method basis-space iterated Stockholder atoms (BS-ISA), we show that the atomic quadrupole moment of a carbon atom in graphene is essentially zero within the limits of precision of the numerical method used.
  We explain how the experimentally measured atomic quadrupole moment of a graphite sample determined by Whitehouse \& Buckingham likely originated almost entirely from edge dipoles, and we propose a more realistic electrostatic model for finite graphene nanoflakes. 
\end{abstract}

%%%%%%%%%%%%%%%%%%%%%%%%%%%%%%%%%%%%%%%%%%%%%%%%%%%%%%%%%%%%%%%%%%%%%
%% Intro/Results/Discussion 1st Half: Develeopment of model from BS-ISA and validation using PAHs
%%%%%%%%%%%%%%%%%%%%%%%%%%%%%%%%%%%%%%%%%%%%%%%%%%%%%%%%%%%%%%%%%%%%%
% AJM test new
% trial 2
% trial 2.1

Electrostatics plays a fundamental role in a wide range of processes including intermolecular interactions, surface adsorption, and crystallisation \cite{Aina2017,Watanabe2011,Cole2016}.
An accurate description of electrostatics is important in simulating these and other chemical and biological processes using force fields. Most force fields used in the simulation of complex systems treat electrostatics through atomic point charges, but increasingly, force fields are employing atomic multipoles through the quadrupole (and sometimes higher) to provide a more realistic description of the electrostatics outside of the overlap region\cite{Sagui2004,Brooks2009,Shi2013,Lagardre2017,Slipchenko2017,Rackers2021}. 

It is well known that  atomic multipoles are not observable experimentally and that different theoretical approaches can lead to appreciably different values of atomic moments.  
Consequently, one may be led to assume that any reasonable definition of the atomic multipoles would suffice. 
Such a conclusion implies that it is meaningless to make any definitive statement about the atomic multipoles, or for that matter, any other property of an atom-in-a-molecule (AIM).\cite{ParrAN05}
This ambiguity has led many researchers to compare definitions of AIM properties using ideas of usefulness, rapidity of convergence, transferability, and, increasingly, by invoking some kind of minimal nature of the definition \cite{Heidar-Zadeh2018,10.1002/cphc.202000040}. 
In this paper, we propose a framework with which one can make meaningful, well-defined, and useful statements about atomic multipoles. 
We will achieve this by focusing on the atomic (in this context we will use the terms ``atomic'' and ``AIM'' interchangeably) multipoles of carbon atoms in polycyclic aromatic hydrocarbons (PAHs) and graphene.

Graphene and graphitic systems in general are important for a wide range of technological processes including batteries, supercapacitors,  desalination, transistors, medicine, and reinforced cementitious composites\cite{Boretti2018,Mbayachi2021,Janavika2023}.
The many uses of graphene and its derivatives make it an important material for which to establish well-founded and accurate atomistic models for use in simulations.
Here we focus on the electrostatic component. 

From the theoretical point of view there are two main reasons for an interest in electrostatic models for graphene: 
First, there is a well-established experimental measurement of what is purported to be the out-of-plane (20) component of the atomic (traceless) quadrupole moment, \QCsph, of carbon in graphite. 
Due to the weak interactions between the graphene layers in graphite, it is often assumed this value of \QCsph should pertain to graphene as well \cite{Zhao2005,Geldart2008,Kocman2014,Xu2021}. 
Second, {\it a priori} one may expect that as graphene has a single type of atom, it will be less ambiguous for theoretical models to determine properties of the carbon atom within this system. 
While these are reasonable assumptions, they do not appear to be valid. 
Indeed, there is a profusion of seemingly contradictory results for the electrostatic properties and models of graphene and graphite:
\begin{itemize}
    \item Whitehouse \& Buckingham (\WB)\cite{Whitehouse1993} have measured the total quadrupole moment (\Qsph) of a graphite sample and from this deduced the atomic quadrupole moment, \QCsph, to be ${-0.675}$ a.u. by dividing \Qsph by the number of carbon atoms in the sample. This \QCsph value has been since used in several simulations \cite{Zhao2005,Geldart2008,Kocman2014,Xu2021}.
    \item Even though \WB reported an experimental value for \QCsph, in modelling electrostatic interactions involving graphene and graphite, other studies have employed values between ${-1.14}$ \cite{Jenness2010,Jordan2019} and ${+0.79}$ \cite{Hansen1995} a.u. In fact, some simulations have ignored the electrostatic contribution of \QCsph altogether, which is functionally the same as stating ${\QCsph=0}$ \cite{Gordillo2005,Marti2010,Ohba2013}. 
    \item Theoretical analysis shows that in the infinite (flat) sheet limit, the atomic quadrupole moments do not contribute to the electrostatic potential \cite{Spurling1980,Kocman2014}. 
    \item \textit{Real} graphene sheets are neither infinite nor flat, but exhibit undulations \cite{Bhatt2022}, possess an edge, and have defects. Edges give rise to edge-dipoles, which to leading order can be shown to contribute in exactly the same way to electrostatic potential as the \QCsph moments. (We will not be concerned with the undulations and impact of defects in this study.)
\end{itemize}
The last point is made clear by considering a continuum model for a finite, circular graphene flake of radius $r$, which is described in detail in the supporting information (SI).
In the continuum model we smear out both the \QCsph moments and the edge dipoles to model a flake terminated with \ce{-CH} bonds around the perimeter. 
The smeared out quadrupole moments result in a uniform quadrupole moment per unit area $\sigma_{20} = \QCsph/A$, where $A$ is the area associated with each carbon atom (half the unit cell area), and the smeared out edge dipoles lead to a dipole per unit length, $\sigma_{10} = \mu_{10}/L$, where $L$ is the arc-length associated with a terminating \ce{-CCHCH-} group and $\mu_{10}$ is the dipole associated with this group. 
As shown in the SI, the potential $V$ at height $d$ on the central axis of the disk is given by
\begin{align}
    V(d) &= \pi (\sigma_{20} - 2 \sigma_{10})  
     \left[
        \frac{r^2}{(d^2 + r^2)^{3/2}}
    \right].
    \label{eq:model_pot}
\end{align}
Some features of the potential are immediately striking and illustrate the points made above:
\begin{enumerate}
    \item The quadrupole moments and edge dipoles contribute with the same dependence on $r$ and $d$. 
    \item In the limit of $r >> d$, $V \rightarrow \pi (\sigma_{20} - 2 \sigma_{10})/r $. I.e., the potential decays very slowly with disk size, and this slow decay can lead to large finite-size effects \cite{Mulvey_2022,Mulvey2024}. 
    \item In the infinite disk limit, the potential $V \rightarrow 0$, as does the electric field. This explains why some studies \cite{Gordillo2005,Marti2010,Ohba2013} modelling adsorption of molecules on graphene and graphite do not account for the quadrupole moment. 
\end{enumerate}

From the first of the above points we see that the edge dipoles are enmeshed in the net quadrupole moment. 
Given this ambiguity, how can we make well-defined statements about the values of the multipole moments of the carbon atoms in a finite flake of graphene? 
Though \WB were well aware of the possibility that the edge dipoles of their finite samples could have a significant impact on their measurement, they ascribed all of the measured quadrupole moment to the carbon atoms to get the experimental value of $\QCsph = -0.675$ a.u.\ 
This may seem a reasonable approach from a chemical point of view as the $\pi$-electron density of carbon atoms in graphene can be expected to result in a non-zero quadrupole moment. 
While this is undoubtedly true for the primitive Cartesian moment $\zzA{C} = \int \rho^{\rm C}(\rr) z^2 d\rr$, the traceless moment $\QCsph = \int \rho^{\rm C}(\rr) \frac{1}{2}(3 z^2 - r^2) d\rr$ has additional contributions from the in-plane density. In both of these expressions, $\rho^{\rm C}(\rr)$ is that part of the density allocated to a single carbon atom.

The key points of the above discussion are as follows: (1) There is an inherent ambiguity of partitioning a bulk electrostatic moment over atomic contributions; 
(2) Although, the long-range electrostatic potential of graphene vanishes in the limit of an infinite sheet, when one considers real graphene samples or comparable analogs like large PAHs, electrostatic interactions can be important; 
(3) Equation \ref{eq:model_pot} demonstrates that for finite samples one cannot disentangle the contributions from the interior atoms from edge contributions. 

With these points in mind, we would like to reframe the discussion by posing a series of questions. 
If one aims to simulate interactions with graphene samples as they actually are, finite, 
do we continue to model electrostatic interactions with only atomic quadrupole moments, or consign all of the net quadrupole moment to the edge dipoles? 
Alternatively, should we consider something in-between, with both edge dipoles and carbon atom quadrupole moments?
Perhaps more importantly, in what sense and why should we attempt to make such definite statements?
The answer to the last question is paramount in constructing a framework within which atomic multipole moments can be objectively assessed.

One possible route to a numerical analysis of this problem is via a fragment approach wherein we estimate the multipoles of carbon in graphene by calculating the atomic multipoles of increasingly larger hexagonal polycyclic aromatic hydrocarbons. 
These prototypical graphene flakes have been the subject of several studies \cite{Jenness2010,Hesselmann2013,Jordan2019} that used the Gaussian distributed multipole analysis (GDMA) \cite{Stone2005} to calculate the atomic multipoles.
In part, these earlier studies were motivated by a desire to use the multipoles of the PAHs in an extrapolation procedure to estimate (\QCsph) for a carbon atom in graphene. 
The motivation for using the GDMA approach was that it is the most widely used technique for distributed multipoles of, at least in principle, arbitrary rank. GDMA multipoles have been used extensively in structure prediction studies of molecular organic crystals \cite {Francia2021,Fowles2021,Francia2020,Price2022,Price2010} and despite issues with how GDMA handles overlap \cite{Lillestolen2009,Kramer2012}, and its uneven convergence properties \cite{Misquitta2014}, it remains widely used.

In this study we counterpoint the GDMA method with the BS-ISA approach which is a basis-space (BS) implementation \cite{Misquitta2014,Misquitta2018} of the iterated Stockholder atoms (ISA) algorithm developed by Lillestolen \& Wheatley \cite{Lillestolen2008,Lillestolen2009}. 
The ISA, and its BS-ISA implementation, belong to a class of Hirshfeld-type Stockholder partitioning methods \cite{Hirshfeld1977,Heidar-Zadeh2018}, in which the properties of the atoms in the molecule are associated with well-defined, exponentially decaying AIM densities. 
The ISA AIM domains are spherical density partitions which can be thought of as the lowest information loss domains obtained by minimizing the Kullback--Leibler divergence \cite{Kullback1951} while allowing for charge flow between the AIMs on formation of a molecule. 
This has been shown \cite{Misquitta2014} to lead to rapid convergence of the AIM properties; a feature that will be important here.
Furthermore, the ISA solution has been shown to be mathematically unique \cite{Lillestolen2009,Bultinck2009}, though in any implementation of this highly non-linear algorithm, numerical choices can lead to apparently non-unique solutions.

In Table ~\ref{table:moments_on_central_atoms} we report the GDMA and BS-ISA atomic moments of carbon atoms on the central ring of the benzene (\ce{C6H6}), coronene (\ce{C24H12}), circumcoronene (\ce{C54H18}), and dicircumcoronene (\ce{C96H24}) sequence of hexagonal PAHs.
These calculations were carried out using the {\sc CamCASP 7.2} \cite{Misquitta2014,Misquitta2018} code with densities obtained from {\sc Psi4} \cite{Parrish2017,Smith2020,Turney2012} using the asymptotically corrected \cite{Gruning2001} PBE0 density functional \cite{Perdew1996,Adamo1999}, denoted here as PBE0(AC), together with the aug-cc-pVTZ basis \cite{Dunning1989,Kendall1992}.
This theory level offers a good compromise between accuracy and computational efficiency, and is known to result in accurate intermolecular interaction energies using SAPT(DFT) \cite{MisquittaPJS05b}, and also, previous results have demonstrated comparable accuracy between MP2 and PBE0 derived properties for PAHs \cite{Mulvey_2022}.
Additional numerical details, including multipole moments on other key atoms are provided in the SI. 

% Trying to get the table to use minus signs instead of hyphens and also align on the decimal.
% It appears to be too long at present.

% needs two packages: dcolumn and array

% align on . and use three decimal places
\newcolumntype{d}{D{.}{.}{1.3}}
\newcolumntype{Z}{>{$}c<{$}} % math-mode for columns

\begin{table}[H]
    \centering
\begin{threeparttable}    
    \begin{tabular}{c  d d d d d  d d d d d  d}
        \hline
        \multicolumn{1}{c}{PAH} &\multicolumn{5}{|c}{\text{BS-ISA}\tnote{1}}  & \multicolumn{5}{|c}{\text{GDMA}\tnote{1}} & \multicolumn{1}{|c}{\text{\WB-like}\tnote{2}} \\ 
        \hline 
        &  \multicolumn{1}{|c}{$\mathrm{q^{\rm C}}$}
           & \multicolumn{1}{c}{$\|\mu^{\rm C}\|$}
              & \multicolumn{1}{c}{$\mathrm{\QCsph}$} 
                 & \multicolumn{1}{c}{$\mathrm{\| \Osph \|}$}
                    & \multicolumn{1}{c}{$\mathrm{\| \Hsph} \|$}
                       & \multicolumn{1}{|c}{$\mathrm{q^{\rm C}}$}
                          & \multicolumn{1}{c}{$\|\mu^{\rm C}\|$}
                             & \multicolumn{1}{c}{$\mathrm{\QCsph}$} 
                                & \multicolumn{1}{c}{$\mathrm{\| \Osph \|}$}
                                   & \multicolumn{1}{c}{$\mathrm{\| \Hsph} \|$}
                                      &  \multicolumn{1}{|c}{$\mathrm{\QCsph}$} \\ 
                                      
        \hline
                                      
         $\mathrm{C_{6}H_{6}}$          &  -0.126       & 0.056     &  0.007  & 0.540  &  0.719          & -0.093        & 0.120     & -1.137  & 1.849 &  2.035        &   -1.057  \\
         $\mathrm{C_{24}H_{12}}$        &  -0.001       & 0.005     & -0.019  & 0.465  &  0.262          & -0.008        & 0.017     & -1.158  & 1.177   &  2.224      &   -0.863  \\
         $\mathrm{C_{54}H_{18}}$        &   0.001       & 0.003     & -0.006  & 0.441  &  0.315          & -0.001        & 0.002     & -1.168  & 1.144  &   2.452      &   -0.812  \\
         $\mathrm{C_{96}H_{24}}$        &  -0.001       & 0.001     & -0.006  & 0.441  &  0.291          & -0.001        & 0.001     & -1.169  & 1.140  & 2.491        &   -0.764  \\
         \hline
        \end{tabular}
        \caption{Average atomic moments of the central carbon atoms in the $\mathrm{C}_{6n^2}\mathrm{H}_{6n} \ n=1-4$ PAHs. Magnitudes are reported for the dipole, octupole, and hexadecapole moments. 
        All moments are in atomic units.}
        \label{table:moments_on_central_atoms}
\begin{tablenotes}\footnotesize
\item[1] The symbols: $\mathrm{q}$, $\|\mu\|$, $\|\mathrm{O}\|$, and $\|\mathrm{H}\|$ in the header row represent atomic charge and the magnitude of the atomic dipole, octupole, and hexadecapole respectively.
\item[2] \QCsph in the ``\WB-like'' column are obtained from the total molecular \Qsph moment by assuming a uniform allocation to all carbon atoms in the corresponding PAH.
\end{tablenotes}\normalsize
\end{threeparttable}
\end{table}

From Table ~\ref{table:moments_on_central_atoms} we see that the GDMA and BS-ISA approaches yield diametrically opposed multipole models for finite graphene flakes.
In the GDMA approach the innermost carbon atoms are described with large atomic multipole moments above the dipole (e.g. $\QCsph < -1$ a.u.). 
In contrast, in the BS-ISA description we have essentially zero \QCsph moments and multipoles above the quadrupole are significantly smaller than those from the GDMA.
The octupole moment of BS-ISA is nearly three times smaller than that of GDMA, and the hexadecapole moment is nearly an order of magnitude smaller.

These large differences in the higher order moments on the interior carbon atoms are accompanied by correspondingly large differences in the charges and dipoles at the periphery of the PAHs.
If one refers to Tables S2 and S3 as well as Figure S4 in the SI, then they will find that all C atoms around the periphery of \ce{C96H24} are negatively charged (whether or not there is a H atom attached) when treated with GDMA, while for the BS-ISA moments the peripheral C atoms without the H atoms are positively charged.
If one combines the atomic charges and dipoles to construct bond dipoles for the terminal C-H moieties, it yields vectors that point towards the center-of-mass (COM) in GDMA and reduce the magnitude of the molecular quadrupole.
Conversely, the edge C-H dipoles constructed from BS-ISA point away from the COM and increase the magnitude of the molecular quadrupole.
Notably, the magnitude of the edge C-H bond dipoles constructed from BS-ISA are $\sim$ 3 to 4 times that of GDMA. The average values range from 0.251 (\ce{C6H6}) to 0.429 a.u. (\ce{C96H24}) for BS-ISA and 0.059 (\ce{C6H6}) to 0.134 a.u.  (\ce{C96H24}) for GDMA.
Note that \textit{both methods yield the same molecular multipole moments when terminated at the same atomic multipolar rank}. 

\begin{figure}[H]
    \centering
    \includegraphics[width=\linewidth]{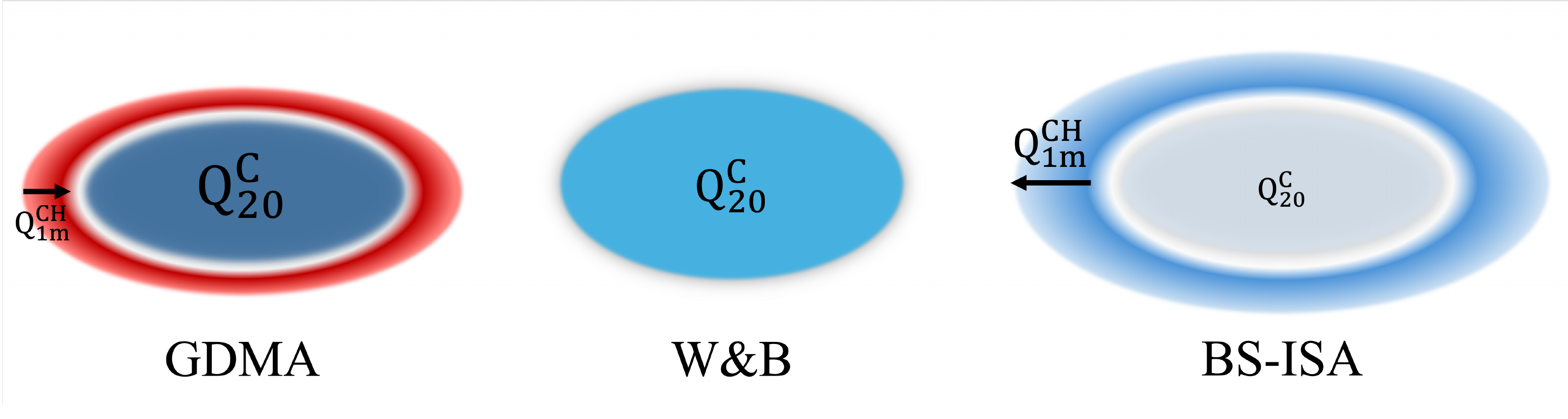}
    \caption{Three different physical models for the electrostatic moments of finite graphene flakes. In the GDMA model the flake possesses a large contribution from the \QCsph moment on the carbon atoms and inward pointing edge dipoles, while in the BS-ISA model there are near-zero \QCsph moments and outward pointing edge dipoles. The \WB model is intermediate with no edge dipoles and only \QCsph moments on the carbon atoms. Intensity of coloration and size of text both reflect the magnitude of the quantity it represents. Red coloration indicates a positive and blue a negative contribution to the total quadrupole (\Qsph).}
    \label{figure:models_for_flakes}
\end{figure}

To these we can include the \WB model as a third distinct physical model in which there are no edge dipoles, and the total quadrupole moment (\Qsph) is the sum of the carbon atom's \QCsph, which is assumed to be the same for all atoms. 
These three models are illustrated in Fig.~\ref{figure:models_for_flakes}. 
We are left with the question of how are we to choose among these?

From the chemical point of view, we expect any system with terminal \ce{-CH} bonds to have outward (from \ce{C} to \ce{H}) pointing dipoles. 
This is consistent with the BS-ISA model, but not that of GDMA.
While it is important that models respect well-defined chemical concepts, our understanding of these concepts has been known to change over time,
so it would be useful to make comparisons using more quantifiable concepts. 

In the absence of an experimentally verifiable means of comparison it is reasonable to impose \textit{a priori} requirements (in the Kantian sense) on the computational models in order to assess their quality.
While this can never be done in an unambiguous manner, there are well-defined requirements that we can expect from computational models. 
For example, there is wide consensus in the quantum chemistry community that computational models should have a well-defined basis-set limit.
Indeed, this requirement was the prime motivation for the development of the hybrid basis- and real-space GDMA algorithm \cite{Stone2005}.

\begin{figure*}
     \centering
    \includegraphics[width=\linewidth]{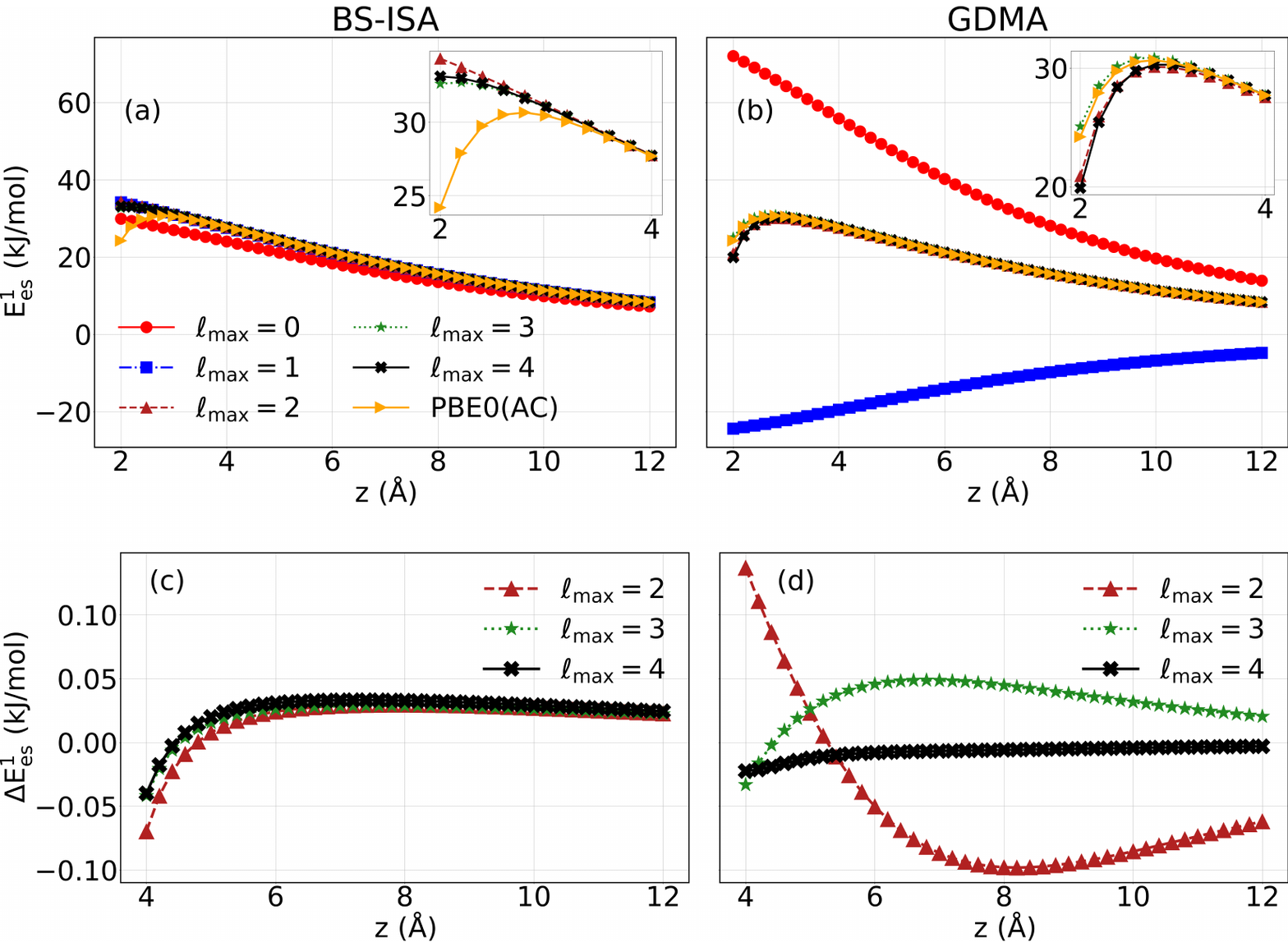}
    \caption{Convergence of multipolar electrostatic interaction energy when a negative point charge is scanned along the principle rotation axis of dicircumcoronene. Results are reported in (a) and (b) for multipole expansions of different orders as well as for the reference results from the PBE0(AC) density. The maximum order of the expansion is given by $\mathrm{\ell_{max}}$. The insets show the results at short range on an expanded scale. Differences between the interaction energies from the various multipole expansions and the PBE0(AC) results are reported in (c) and (d).}
    \label{figure:c96h24_scan}
\end{figure*}

In this work we require that the ``best'' model be as simple as possible while remaining accurate. 
Figure~\ref{figure:c96h24_scan} reports the electrostatic energy of a negative point charge interacting with \ce{C96H24} when it is scanned along the principle rotational axis of the molecule. 
Results are reported for the reference PBE0(AC) charge density and for the GDMA, and BS-ISA multipole expansions at successively higher order atomic multipolar ranks.  
Additional details for these calculations are provided in the SI document.

Figures \ref{figure:c96h24_scan} (a) and (b) presents the interaction energies obtained from the various GDMA and BS-ISA multipole expansions and the reference PBE0(AC) results for point charge-PAH separations ($z$) between $2$ and $12$ \AA. 
The turnover starting at $\mathrm{z} \approx  4$ \AA \ in the electrostatic potential from the reference calculations is due to charge penetration, which is absent from the electrostatic potentials of the multipole expansions.
Figures \ref{figure:c96h24_scan} (c) and (d) reports the differences between the electrostatic interaction energies from the various multipole expansions truncated at maximum rank $\ell_{\rm max}$, $\mathrm{E^{1}_{es}(\ell_{max})}$, and the reference, non-expanded electrostatic energy from the PBE0(AC) densities, $\mathrm{E^{1}_{es}(PBE0(AC))}$.
This difference, $\mathrm{\Delta E^{1}_{es}}(\ell_{\rm max}) = \mathrm{E^{1}_{es}(PBE0(AC))} - \mathrm{E^{1}_{es}(\ell_{max})}$, gives us a working definition of the charge-penetration energy for a multipole expansion of maximum rank $\ell_{\rm max}$. 
From these figures, it is seen that at distances where charge penetration is negligible, the electrostatic potential from the BS-ISA calculations is already well converged by atomic charges, dipoles, and quadrupoles ($\mathrm{\ell_{max}}=2$). 
In fact, BS-ISA, even when truncated to include only atomic charges, gives a reasonably accurate representation of the electrostatic interaction energy.
In contrast, with the GDMA procedure it is necessary to include moments through the hexadecapole $\mathrm{\ell_{max}}=4$ to obtain well converged results. 
The electrostatic potentials from the $\mathrm{\ell_{max}}=2$ and $\mathrm{\ell_{max}}=3$ GDMA expansions differ significantly from the reference result.
The very small error in the BS-ISA  electrostatic potential over this range of distances, when including terms through $\mathrm{\ell_{max}}=4$, is expected to be due to the errors introduced by use of density fitting basis sets.  

As already noted, the electrostatic penetration energy comes into play at distances shorter than $4$ \AA{} for the systems of interest.
A host of theoretical approaches have been proposed for modeling charge penetration \cite{Wheatley1993,Gordon2000,Piquemal2003,Piquemal2006,Cisneros2006,Elking2010,Rackers2015,Rackers2017}. 
We adopt a pragmatic, operational definition of charge penetration energy: It is a contribution to electrostatics due to the interpenetration of separate charge distributions, for which a classical point multipolar expansion fails to account.
This definition bears similarity to that employed by Stone, for which he demonstrates analytically that the charge penetration energy should decay exponentially with distance between the charge distributions\cite{Stone2013}, which is concomitant with the exponential decay of electronic density.

\begin{figure*}
    \centering
    \includegraphics[width=\linewidth]{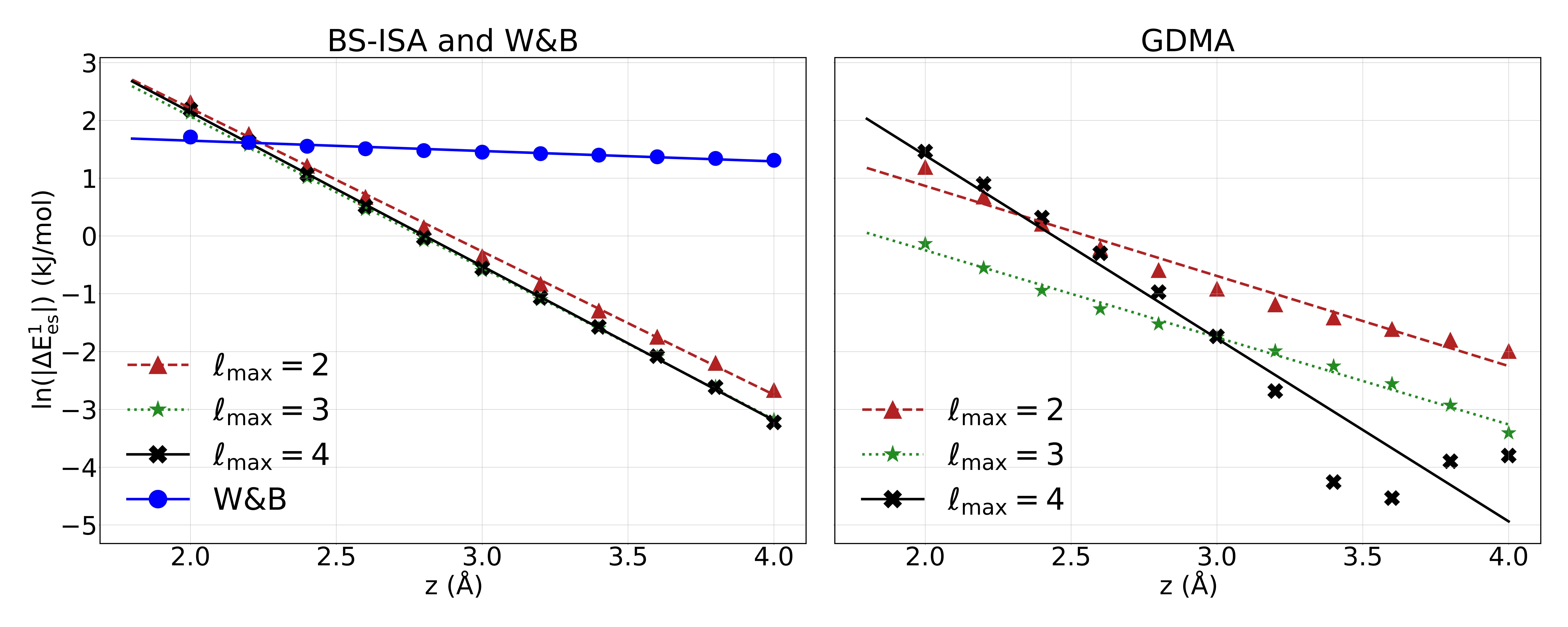}
    \caption{Convergence of charge penetration energy in multipolar electrostatic models when a negative point charge is scanned along the principle rotation axis of dicircumcoronene. The maximum order of the expansion is given by $\mathrm{\ell_{max}}$. The figure shows the natural log of the magnitude of energy differences reported in Figures \ref{figure:c96h24_scan} (c) and (d). The left image we include the charge penetration energy if one applies a \WB-like approach by simply allocating the molecular quadrupole moment over all carbon atoms equally as \QCsph. }
    \label{figure:c96h24_scan_c}
\end{figure*}

The intent of this work is not to develop another theory of charge penetration, but we do believe it is instructive to compare the short-range behavior of the three electrostatic models considered thus far.
To ascertain differences in charge penetration energy in the three models we take the natural log of the energy differences shown in Figure~\ref{figure:c96h24_scan_c} for $2 \leq z \leq 4$ \ \AA.

There are two criteria guiding us in this assessment.
First: if the electrostatic model is physically sensible at short range, then one should observe a linearly decreasing trend in the natural log of charge penetration with increasing $z$\cite{Stone2013}.
Second: A practically useful electrostatic model should describe charge penetration energy with a minimal order expansion, i.e. it should be relatively well-converged at low-rank. 

As seen in Figure \ref{figure:c96h24_scan_c}, the deviation of the electrostatic potentials from the BS-ISA  multipole expansions from that of PBE0(AC) displays the expected exponential distance dependence at short range, with nearly the same results whether the expansion is truncated at $\mathrm{\ell_{max}}=2,3, \ \mathrm{or} \ 4$. 
The situation is very different in the case of the GDMA expansion. 
Not only is the sign of this energy difference incorrect for the $\mathrm{\ell_{max}}=2$ and 4 expansions (see inset of Figures \ref{figure:c96h24_scan} (a) and (b) ``GDMA''), the GDMA electrostatic potential over this distance range depends strongly on the order of the expansion. 
The behavior of the \WB-like model at short range is questionable as well, as the difference from the PBE0(AC) electrostatic potential decays much more slowly with $z$ than the other two models. 
It is worth noting that regardless of short-range behavior the penetration energies of all electrostatic models, including the \WB-like model, tend to zero at large values of $z$.
A figure demonstrating this is included in the SI document. 

Figures \ref{figure:c96h24_scan} and \ref{figure:c96h24_scan_c} demonstrate features that favour the BS-ISA atomic multipoles.
At distances where charge penetration is relatively unimportant, the BS-ISA multiple expansion converges much more rapidly than the GDMA expansion, and at distances where charge penetration is important BS-ISA, unlike GDMA or the \WB-like model, displays physically correct behavior. 
This analysis lends support to the conclusion that the distributed multipole moments from BS-ISA reflect an underlying physical reality more closely than those from GDMA or the \WB-like approaches. 

One might expect that they could establish the \QCsph of a carbon atom of graphene definitively from an electronic structure calculation on graphene using periodic boundary conditions. 
The idea here being to calculate the quadrupole moment of the two-carbon atom unit cell and then dividing by two to obtain a value of the moment of a single carbon atom. 
However, due to the use of periodic boundary conditions, only the primitive (i.e. containing its trace) quadrupole moment $\zzA{cell}$ can be uniquely determined from such a calculation, as the in-plane components, $\xxA{cell}$ and $\yyA{cell}$, which are equal by symmetry, become origin-dependent \cite{Wheeler2019}. 
In spite of this limitation, it is instructive to calculate $\zzA{cell}$ for a unit cell of  the graphene sheet, from which, $\zzA{C} = \zzA{cell}/2$, the primitive moment for a carbon atom in the cell, can be determined.

To accomplish this, DFT calculations have been carried out for graphene with periodic boundary conditions, employing a two-atom cell with up to 16 \AA{} of separation between the graphene layers in the aperiodic direction. 
The CASTEP\cite{Castep} code was used, and the LDA \cite{Kohn1965}, PW91 \cite{Burke1998}, PBE \cite{Perdew1996_pbe,Perdew1997}, and PBE0 \cite{Perdew1996,Adamo1999} functionals were tested.
Both norm-conserving\cite{Hamann1979} and ultra-soft\cite{Vanderbilt1990} pseudopotentials were tested.
Additional details on the calculations are provided in the SI document.

The tested GGA functionals yielded primitive carbon quadrupole moments $\zzA{C}$ within a narrow range of values from $-3.922$ a.u.\ to $-3.951$ a.u., and the hybrid functional, PBE0, with a norm-conserving pseudopotential\cite{Hamann1979} yielded $\zzA{C} = -3.931$ a.u. 
These results are close to the $-4.027$ a.u.\ value of $\zzA{C}$ determined using the BS-ISA procedure for the C atoms of the central ring of \ce{C96H24}, which further supports the BS-ISA model.

In our discussion of the \WB determination of \QCsph of graphite, we highlighted the problem posed by edge dipoles. 
While this is a non-issue for idealized, infinite graphene sheets, experimental samples are necessarily finite and will contain edge effects, the impact of which should not be neglected.
To illustrate this further, we constructed two simple electrostatic models of graphene nanoflakes using geometrical parameters that are identical to those employed for the PAHs we have considered thus far except that the hydrogen atoms have been removed (i.e., $\mathrm{C}_{6n^2}, \ n=2,3,\dots$). 

In the first of the two models we place a \QCsph moment on every carbon atom in a nanoflake.
For the \ce{C24}, \ce{C54}, and \ce{C96} nanoflakes we use the BS-ISA values of \QCsph tabulated in Table \ref{table:moments_on_central_atoms}, and for nanoflakes larger than \ce{C96} we use the \QCsph value of the central carbon atoms of \ce{C96H24} as this quantity has converged with respect to system size and should change little for larger nanoflakes (again refer to Table \ref{table:moments_on_central_atoms}). 

\begin{figure}[H]
    \centering
    \includegraphics[width=\linewidth]{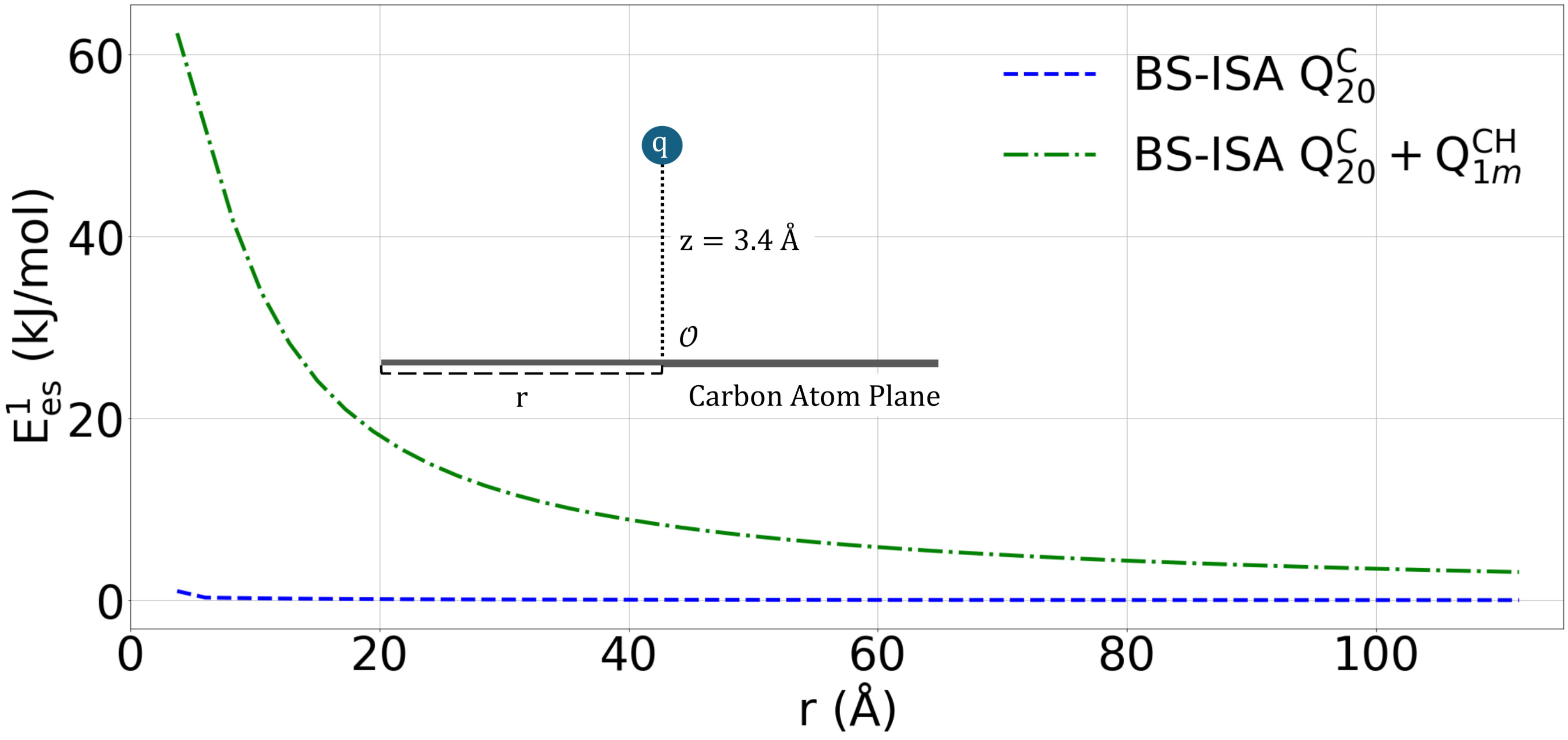}
    \caption{The electrostatic interaction of a negative point charge (q) with increasingly large carbon nanoflakes at a distance of $\mathrm{z}=3.4$   \AA \ above the plane of the nanoflake. Results shown for only atomic quadrupole moments ($Q_{20}^C$) on the carbon atoms as well as those with  dipoles ($Q^{\mathrm{CH}}_{1m}$) on the edge carbon atoms. The axis, $r$, is an average distance of the edge carbons to the center of mass in the carbon nanoflake. A schematic of the arrangement is included as an inset for clarity.}
    \label{figure:model_nanoflake_scan}
\end{figure}

In the second model, we also include edge dipoles ($Q^{\mathrm{CH}}_{1m}$) to the terminal carbon atoms of the nanoflakes.
This crudely models the electrostatics of a PAH, as the orientation and magnitude of the edge dipoles were informed by the edge C-H dipoles of coronene, circumcoronene, and dicircumcoronene as described by BS-ISA.
The edge dipoles for nanoflakes larger than dicircumcoronene were obtained via a fitting procedure detailed in the SI document.
Figure \ref{figure:model_nanoflake_scan} shows the electrostatic interaction of a negative point charge with increasingly large graphene nanoflakes represented via these two electrostatic models.

As seen from the figure, the contribution of the atomic quadrupoles is unsurprisingly negligible over the entire range of nanoflakes considered. 
On the other hand, the contribution of the edge dipoles to the electrostatics is orders of magnitude larger, being $\sim 3.1$ kJ/mol for the largest system considered ($15000$ C atoms) with a radius of about $111.5$ \AA.
Although the graphitic nanoflakes considered by \WB had a much larger radius than those considered here and their samples included contributions from multiple graphene layers, the above analysis illustrates how edge effects can dominate the net electrostatic interaction. 

%%%%%%%%%%%%%%%%%%%%%%%%%%%%%%%%%%%%%%%%%%%%%%%%%%%%%%%%%%%%%%%%%%%%%
%% Conclusions 
%%%%%%%%%%%%%%%%%%%%%%%%%%%%%%%%%%%%%%%%%%%%%%%%%%%%%%%%%%%%%%%%%%%%%

In this article we have presented results from the distributed multipole methods GDMA and BS-ISA for the atomic multipoles of increasingly large PAHs.
Using the atomic multipoles of the innermost carbon atoms of the PAHs we were able to estimate the quadrupole moment of the carbon atoms of an idealized infinite graphene sheet.
The resulting estimates of $\QCsph$ of a carbon atom in graphene represent two divergent results, $-0.006$ a.u. for BS-ISA and $-1.169$ a.u. for GDMA, between which lies a third, the experimental measurement by Whitehouse and Buckingham, $-0.675$ a.u.
These results may seem incompatible when one considers the total quadrupole moment, but as demonstrated above they are, in fact, consistent when one accounts for the contribution of the edge multipoles as depicted in Fig.~\ref{figure:models_for_flakes}.
As explained above, the contribution of an atomic quadrupole is immaterial to the total electrostatic potential in an infinite system, but this limit is attained very slowly, as seen in Fig.~\ref{figure:model_nanoflake_scan}.
All physical samples are finite systems with edges --- and this includes the experimental setup of Whitehouse and Buckingham --- which necessitates that any realistic model account for the contribution of the edge electrostatic moments.
This rules out models neglecting edge effects when dealing with finite systems, but leaves one to choose between the BS-ISA and GDMA models.

The dissection of the total electrostatic moments into atomic multipoles is not unique and cannot be established experimentally, so arguments in favour of one model over another must come from imposing criteria and desired features.
In the absence of external benchmark data, we adopt criteria that select for a minimal model: Accurate at long-range, low complexity in its representation (i.e., rapidly convergent with multipolar rank), and yields the expected exponential behavior of short-range charge penetration energy.
At a sufficiently large expansion of the multipolar series, the hexadecapole, both GDMA and BS-ISA agree in their description of long-range electrostatics.
However, our criterion of a rapidly convergent multipolar expansion separates the two models at long-range as GDMA requires up to the atomic octupole to achieve the same accuracy as BS-ISA terminated at the atomic dipole.
Furthermore, the BS-ISA model is physically sensible at short-range producing an exponentially decaying charge penetration contribution to the electrostatic energy that is nearly insensitive to maximum rank of the multipolar expansion, whereas the error using the GDMA model depends strongly on the order of the expansion.

Given these observations, we conclude that the BS-ISA model provides a more realistic description of the electrostatics of large, but finite graphene analogues. 
Within this model, the total quadrupole of a finite graphene flake arises almost entirely from permanent electrostatic moments at the edges (dipole moments to leading order). 
To illustrate the impact of these edge moments, we calculated the electrostatic potential of graphene nanoflakes containing up to $15000$ C atoms using atomic moments derived from the BS-ISA model.
Fig.~\ref{figure:model_nanoflake_scan} shows that the contribution of the atomic quadrupoles from BS-ISA negligible over the entire range of nanoflakes considered, but the contribution of the edge dipoles is orders of magnitude larger, even at the center of a graphene flake where the edge dipoles are a distance of $\sim 111.5$ \AA \ away.
Thus, for any realistic simulations of finite graphene flakes, the electrostatic error will only grow larger as one approaches the edges of the carbon sheet unless these electrostatic features are present.

Although our focus has been on hexagonal PAHs, our conclusions are also relevant for graphene nanoflakes with edge terminations other than \ce{CH} groups. 
With the above in mind, we assert that for flat graphene nanoflakes, the atomic quadrupole of carbon is immaterial for intermediate to long-range intermolecular interactions and more realistic force-field simulations incorporate the relevant effects of edge termination into electrostatics.

\begin{acknowledgement}

DM acknowledges support from US National Science Foundation (Award No. 2142874) and the Camille and Henry Dreyfus Foundation (Award No. TH-23-033) for support of this research. Both DM and KDJ acknowledge support from the US National Science Foundation under grant number CBET2028826. AJM acknowledges support from UKRI grant number EP/X036863/1. The authors acknowledge Dr Michael J. Rutter for performing the PBC electrostatic calculations in CASTEP. Some calculations were carried out on computers in the University of Pittsburgh’s Center for Research Computing and Data. 
\end{acknowledgement}

\begin{suppinfo}

Details describing the methodology used for obtaining multipole moments from BS-ISA, GDMA, and periodic boundary condition calculations, the evaluation of electrostatic interaction and charge penetration scans, and the process for constructing or model edge dipoles are available at \textcolor{red}{Link to Supporting Information Document}. Geometries, example inputs and outputs, plotting scripts, and more available at \url{https://github.com/dev-m-mulvey/quadrupole.git}

\end{suppinfo}

\bibliography{main}

\end{document}

%% file: macros.tex
\newcommand{\Qsph}[0]{\ensuremath{\mathrm{Q}_{20}}\xspace}

\newcommand{\Osph}[0]{\ensuremath{\mathrm{O}^\text{C}}\xspace}
\newcommand{\Hsph}[0]{\ensuremath{\mathrm{H}^\text{C}}\xspace}
\newcommand{\QCsph}[0]{\ensuremath{\mathrm{Q}_{20}^\text{C}}\xspace}
\newcommand{\zzA}[1]{\ensuremath{\langle zz \rangle^{\text{#1}}}\xspace}
\newcommand{\yyA}[1]{\ensuremath{\langle yy \rangle^{\text{#1}}}\xspace}
\newcommand{\xxA}[1]{\ensuremath{\langle xx \rangle^{\text{#1}}}\xspace}

\newcommand{\WB}[0]{W\&B\xspace}

\newcommand{\rr}[0]{\ensuremath{{\mathbf r}}}